\documentclass[11pt, a4paper]{article}
\usepackage{psvectorian}
\usepackage{graphicx}
\usepackage{amssymb}
\usepackage{enumitem}

\usepackage{geometry}
\usepackage{tikz}
\usepackage{float}
\usepackage{url}
\usepackage{mathtools}
\usepackage{amsmath,amsfonts,amsthm}
\usepackage{mathrsfs, amsmath}

\newtheorem{question}{Question}
\newtheorem{solution}{Solution}
\usepackage{subfig}

\usepackage{subfig}
\usepackage[colorlinks,
linkcolor=blue,
anchorcolor=blue,
citecolor=blue
]{hyperref}
\begin{document}
\begin{center}
{\large \bf The Effect of Using Popular Mathematical Puzzles on The Mathematical Thinking of Syrian Schoolchildren}
\end{center}
\begin{center}
Duaa Abdullah$^{1*}$ \&  Jasem Hamoud$^{2*}$\\[6pt]
  $^{*}$ Physics and Technology School of Applied Mathematics and Informatics \\
Moscow Institute of Physics and Technology, 141701, Moscow region, Russia\\[6pt]
Email: $^{1}${\tt duaa1992abdullah@gmail.com}, $^{2}${\tt hamoud.math@gmail.com}.
\end{center}
\noindent
\begin{abstract}
In this paper we provide a good overview of the problems and the background of mathematics education in Syrian schools. We aimed to study the effect of using popular mathematical puzzles on the mathematical thinking of schoolchildren, by conducting a paired experimental study (pre-test and post-test control group design) of the data we obtained through a sample taken from students of sixth-grade primary school students in Syria the Lady Mary School in Syria, in order to evaluate the extent of the impact of popular mathematical puzzles on students' ability to solve problems and mathematical skills, and then the skills were measured and the results were analyzed using a t-test as a tool for statistical analysis.
\end{abstract}

\noindent\textbf{AMS Classification 2010:} 397C70, 97U40, 97F90 , 97D60, 97D70, 97A40.


\noindent\textbf{UDC:} 37.013,51.

\section{Introduction}\label{sec1}

Puzzles and games are ancient tools for mental challenges, enhancing critical thinking skills and engaging students in mathematical exploration. Popular puzzles are an effective tool for improving problem-solving skills in mathematics education. Solving puzzles requires logical thinking, pattern recognition, and the ability to apply mathematical concepts in creative ways. By incorporating puzzles into mathematics education, students can develop these skills and increase their confidence in problem-solving. Puzzles also add excitement and fun to mathematics, helping students maintain their enthusiasm and interest in the subject.
Learning enables our students to develop positive attitudes toward mathematics, and even the weakest students gradually become involved in solving projects~\cite{Kubinova}.

What we have learned is that we, as humans, do not only operate with the logical and analytical sides of our brains.
In fact, what experts call "inferring meaning" plays a crucial role in analytical thinking. This process can make solving puzzles extremely beneficial. As the study continued, it became clear that people don't just use the analytical side of their brains to understand and replicate a pattern.
It also clearly demonstrated that the skill of understanding meaning is very important and can be developed when the mind is pushed to do something challenging, enjoyable, and thus rewarding.

These are the key words psychology uses to describe the fertile ground available for effective learning: challenging, but not frustratingly impossible, and enjoyable. This is precisely the kind of resource that puzzles provide for mathematics; they are important, as mathematics is often viewed as a boring, overly academic subject with no real-life relevance. I hope to explore how puzzles can be a tool for enhancing mathematics education for all students, from primary to advanced levels.

In this paper, we discuss and explain the difficulties in teaching and learning mathematics in Syrian schools. It identifies and analyzes the causes of students' poor performance in mathematics learning and academic achievement.
We believe this paper is one of the few documents that discusses mathematics education in Syria. In the current context, due to the dire situation that Syria has been experiencing for more than thirteen years in a state of war, the country is now exhausted and in need of any opportunity to recover, especially with regard to improving the level of education. 
%---------------------------------

\section{Research Problem and Study Objective}\label{sec2}

The educational reality now in Syria is in dire need of building and rising again, after researchers noticed that many primary school students face difficulty in learning mathematics and have weak mental arithmetic ability. Because the educational process is limited to theoretical material devoid of application, the educational reality now in Syria is in dire need of building and rising again, after researchers noticed that many primary school students face difficulty in learning mathematics Weak mental arithmetic ability; because the educational process is limited to theoretical material devoid of application and relies on ordinary methods devoid of excitement, and since modern educational trends focus on moving away from indoctrination in teaching mathematics and activating the student’s role in the educational process. So, this study aimed to choose a method of teaching mathematics based on solving common puzzles through play. This study came to investigate and evaluate the effect of using common puzzles on mathematical thinking and their role in developing mental arithmetic skills among sixth-grade students. Specifically, we have tried to answer the following questions:
\begin{enumerate}
	\item Is there a statistically significant difference at ($\alpha=0.05$) between the average scores of sixth-grade students in both the control group (which learned in the traditional way) and the experimental group (which learned using common puzzles) on test of natural numbers and integers, attributed to the method of teaching with common puzzles so that we can notice the difference in mathematical thinking?
	\item Is there a statistically significant difference at ($\alpha=0.05$) between the average scores of students on test of natural numbers and integers in terms of mathematical thinking and mental arithmetic ability by comparing the experimental and control groups?
\end{enumerate}

%---------------------------------
\section{Research Importance} \label{sec3}

This paper addresses a critical issue-mathematics education in a crisis-affected country (Syria). There is a growing interest in understanding and improving education in such contexts, making the paper highly relevant to contemporary educational discourse. This study focuses on sixth-grade primary school students, a crucial stage for developing foundational mathematical skills and attitudes towards the subject. The use of puzzles and games is an alternative method to assess students’ level of mathematical thinking; it enhances what teachers tend to know about students’ understanding to help them overcoming their misconceptions and measuring their performance in a real situation.

This study is considered as one of the few studies in the Arab world that examines the effect of using puzzles and games in improving students’ mathematical thinking, attributable to their self-efficacy.

\section{Research Hypotheses}\label{sec4}

The present study sought to test the following hypotheses:
\begin{enumerate}
	\item The mean scores of the experimental and the control groups in the mathematical thinking test are not statistically different from each other at $(\alpha=0.05)$.
	\item The interaction between the group (experimental, control) and self-efficacy (high, moderate) does not have statistically significant differences $(\alpha=0.05)$ of the mean such that the effect of self-efficacy is aggregated on the mathematical thinking test.
\end{enumerate}

\subsection{Study Population and Sample}\label{sec4.1}

The study population consisted of all sixth-grade primary school students at Our Lady of Mary School in Syria, according to their level of effectiveness, by conducting an associated exploratory study on them during the first semester of the 2023-2024 academic year, and their number is about (400) students.

The study sample consisted of (83) male and female students in the basic education stage, sixth grade, in mathematics, distributed into two groups. The first group, numbering (42) students, was randomly selected as an experimental group that learned the natural numbers and integers using puzzles and games, and the other group, numbering (41) students were taught in the traditional way, and the researchers taught the two groups equally together.

\subsection{Study limitations}\label{sec4.2}

The boundaries in the study were divided into:
\begin{itemize}
    \item \textbf{ Objective limit:} This study was limited to all chapters related to natural numbers (addition, subtraction, division, and multiplication) in the first and second units of the mathematics textbook for the sixth grade of primary school, in addition to integers.
    \item \textbf{Human limit:} Sixth grade students.
    \item \textbf{Time limit:} first semester of the academic year (2023-2024).
    \item \textbf{Spatial limit:} The School of the Virgin Mary in Aleppo, Syria.
\end{itemize}

\section{The most Common Problems in Syrian’s Schools}\label{sec5}

Today, after more than thirteen years of continuous experience in the field of education in Syria, whether in private or government schools, or in charitable educational associations and institutions that contributed to supporting education in Syrian schools, especially during the war period, the war in Syria has had a devastating impact on the education system in the country, which left many problems and challenges in the education sector in Syria We can summarize the most common problems in Syrian schools as follows:
\begin{enumerate}
    \item Lack of educational resources and means: Many schools were damaged or destroyed during the conflict. This means a shortage of classrooms, textbooks, educational tools or any sports laboratories to assist in the educational process, and there may also be limited access to basic necessities such as clean water and sanitation.
    \item Teacher shortage: The war led to the displacement of many teachers, leaving schools suffering from a shortage of staff. This can lead to overcrowded classrooms and lack of individual attention to students.
    \item Displaced students: Millions of Syrians have been internally displaced or have become refugees. This disrupts children's education and makes it difficult for them to enroll in new schools, especially if they lack documents or suffer from a different Arabic dialect.
    \item Limited curricula: The ongoing conflict may limit the possibility of updating the curricula, as the mathematics curriculum is traditional and huge, often based on many theoretical topics far from the possibility of practical application of each of them, and must be completed in a short time, and this causes aversion to the direction of mathematics.
    \item Weak financial support for teachers: This is the main reason that weakens teachers’ motivation to develop their teaching skills, their dedication to their work and performing it honestly.
    \item Poor professional performance of the teacher: teaching without an objective will not show the relevance of the material in working life, how it is used, and its applied usefulness (which is the greatest motivation for learning).
    \item Absence of a strong mathematical foundation: weakness in performing the mathematical operations required in the learner’s daily life, especially these days due to the high prices and high cost of living.
    \item Psychological trauma and psychosocial problems: Children who have lived through war may suffer from psychological trauma and psychological distress. This may make it difficult for them to concentrate at school.
    \item Security concerns: Security threats and ongoing violence may pose a risk to children as they travel to and from school.
\end{enumerate}

It is important to note that the situation varies depending on the geographical location within Syria. However, these represent some of the wide-ranging challenges facing the Syrian education system~\cite{Borovik}.
%--------------------------------------------
\subsection{Background of Mathematics Education in Syria}\label{sec5.1}

Mathematics education has a particularly distinguished history in Syria. From ancient times till the present, mathematicians belonging to this land left their mark on the world's mathematics. Greek mathematician Zenon, known for his paradoxes, was one of them. However, the education system in Syria has always had an effect from foreign invasions, and so has been the case with mathematics education. The issue is further complicated by the dual education system, the model based on the French schooling system at the private and international schools, and the Arabic model used in public schools~\cite{Almustafa}. 
Syria, a country of 18 million people, has come into the list of IMU for having 98\% proficiency in mathematics. These percentages are in conflict when considering the quality of mathematics education available, and may be due to the youth population, extremely high education standards, and some cultural and historic emphasis on mathematics. However, this analysis does infer that there are much worse situations in other countries on this list~\cite{Englert}. 

Math is an integral subject to study in Syria. High school students have to take it or face severe penalties. Kagba adds schools must be in a position to show that some students just cannot take math and there have to be alternatives for them promoting them to take higher level courses in literature and sciences. It is not only a compulsory subject, but it is also a gateway to various higher studies, no matter the field. This adds extra emphasis on needing to improve the current system and its teaching. On the other hand, the Arabic culture, while having a positive stance towards mathematics, does not see it as a problematic subject. This may be a reason as to why there is an astonishing lack of research in mathematics education in Syria~\cite{Djordjevic}.
%-------------------------------------------
\subsection{Importance of Addressing Popular Mathematics Problems} \label{sec5.2}

The use of puzzles in education is not entirely new, this paper focus on ``popular mathematical puzzles'' and their structured application as a primary teaching method for specific mathematical concepts (natural numbers and integers) in the Syrian context is distinctive. The Syrian education system is characterized by a number of problems. Education reformers in Syria argue that changes need to be made in areas of creativity, resources, and attitudes towards education in order to build a generation with enhanced problem-solving abilities. This is an important point. Often, it is easy to confuse excellence in exams with excellence in learning. The former can be achieved through hard work and memorization, but the latter requires a different approach. Excellence in learning can be assessed through the ability to tackle unfamiliar problems and produce elegant solutions, and thus is very difficult to quantify. In spite of the lack of clarity on this issue, there is a general feeling that mathematics education in Syria is not helping students to achieve problem-solving skills, and an investigation of the popular mathematics problems in Syrian schools confirms this~\cite{Grcar}.

Mathematics is a core subject in the Syrian curriculum, with the number of teaching hours allocated to math greater than that for any other subject. This reflects the emphasis the educational system places on mathematics, a reflection of values deeply held by Syrians. These values include respect for learning, belief in the power of education, and faith in its ability to change the future for the better. Education is also a high priority for many Syrian families who are prepared to make considerable sacrifices to ensure a better future for their children; for many families, this means spending a sizeable portion of their income on private tuition~\cite{Jorgensen,Popel}.

%----------------------------------------
\section{The Popular Puzzles} \label{sec6}

The researchers designed and developed the popular puzzles as a media to be used in the learning process. In addition, the researchers assign them to students in the experimental group as a task in harmony with the vocabulary learned on the subject taught, as an assessment tool during the instructional process, and it is a follow-up to the solution that was offered by the student.

Then they talk about the task with the class. The control group received instruction on the same unit traditionally, which is by practicing from tasks available in the textbook solely. It was around one task on each lecture, either a puzzle or a game, and there was a call for students to work along these tasks.

In determining a puzzle in the classroom teaching and learning process, the researcher designed for different levels of the puzzle.

When using a puzzle in the teaching and learning process in the classroom, different levels of the puzzle were planned. Table 1 describes the puzzle of Magic Triangle:
\begin{question}~\label{ques1}
Using this Magic Triangle with natural numbers, fill the circles on the triangle so that each edge sums to 17.
\end{question}
\begin{solution}
In order to solve Question~\ref{ques1}, we have a puzzle. To find a solution to this puzzle, the question must be analyzed carefully. We notice that each edge sums to 17. Then, through Figure~\ref{figen1} we find that
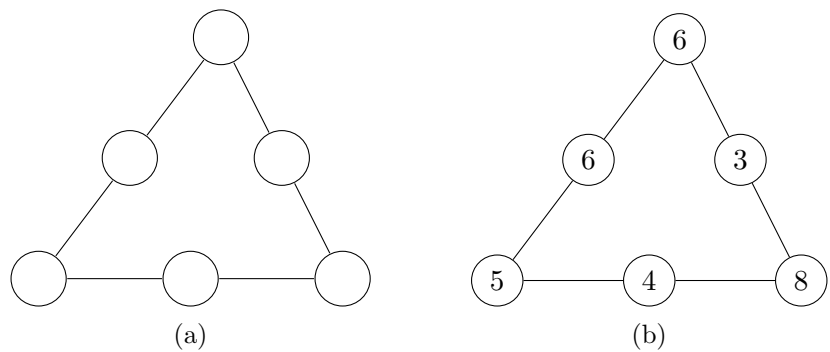
\begin{figure}[H]
  \centering
  \begin{tabular}[b]{c}
   		\begin{tikzpicture}[every node/.style={draw,circle, scale=2}, scale=.4];
        
				\node (a) at (0,0) { };
				\node (b) at (6,8) { };
				\node (c) at (10,0) { };
				\node (d) at (3,4) { };
				\node (e) at (8,4) { };
				\node (f) at (5,0) { };
				\draw (a)--(d)--(b)--(e)--(c)--(f)--(a);
			\end{tikzpicture} \\
    \small (a)
  \end{tabular} \qquad
  \begin{tabular}[b]{c}
   	\begin{tikzpicture}[every node/.style={draw,circle, scale=1}, scale=.4]
				\node (a) at (0,0) { $5$};
				\node (b) at (6,8) { $6$};
				\node (c) at (10,0) { $8$};
				\node (d) at (3,4) { $6$};
				\node (e) at (8,4) {$3$ };
				\node (f) at (5,0) {$4$ };
				\draw (a)--(d)--(b)--(e)--(c)--(f)--(a);
			\end{tikzpicture} \\
    \small (b)
  \end{tabular}
  \caption{Triangle Puzzle (a)~and solution (b).}~\label{figen1}
\end{figure}
\end{solution}

\begin{question}~\label{ques2}
Fill this magic triangle pictured below with integer numbers, (positive and negative) in the solution so that each edge sums to 9.
\end{question}
\begin{solution}
To find a solution to this puzzle, the Question~\ref{ques2} analyzed as we show that by Figure~\ref{figen2}, for each edge sums to 9.
\begin{figure}[H]
  \centering
  \begin{tabular}[b]{c}
   		\begin{tikzpicture}[every node/.style={draw,circle, scale=2}, scale=.4];
				\node (a) at (0,0) { };
				\node (b) at (6,8) { };
				\node (c) at (10,0) { };
				\node (d) at (3,4) { };
				\node (e) at (8,4) { };
				\node (f) at (5,0) { };
				\draw (a)--(d)--(b)--(e)--(c)--(f)--(a);
			\end{tikzpicture} \\
    \small (a)
  \end{tabular} \qquad
  \begin{tabular}[b]{c}
   	\begin{tikzpicture}[every node/.style={draw,circle, scale=.8}, scale=.4]
				\node (a) at (0,0) { $3$};
				\node (b) at (6,8) { $-4$};
				\node (c) at (10,0) { $8$};
				\node (d) at (3,4) { $10$};
				\node (e) at (8,4) {$5$ };
				\node (f) at (5,0) {$-2$ };
				\draw (a)--(d)--(b)--(e)--(c)--(f)--(a);
			\end{tikzpicture} \\
    \small (b)
  \end{tabular}
  \caption{Triangle Puzzle each edge sums to 9 (a)~and solution (b).}~\label{figen2}
\end{figure}
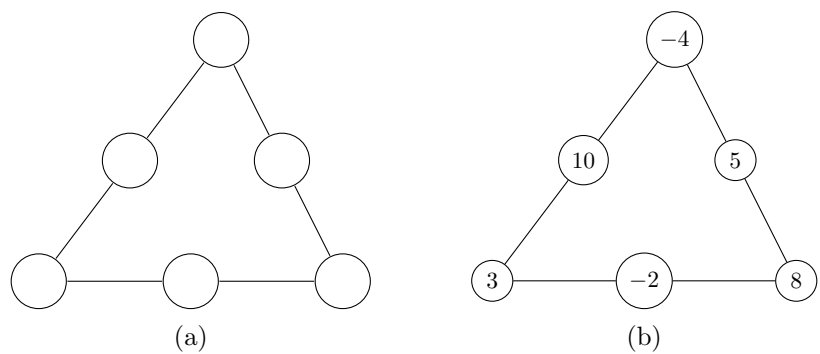
\end{solution}

In Question~\ref{ques1}, the set of natural numbers was determined to fill the triangle, while in Question~\ref{ques2} was developed to a higher level of thinking so that the set of integers is used, so the student enjoyed the puzzle and used addition and subtraction skills to think in a problem-solving manner.

\subsection{Study Procedures}\label{sec6.1}
	
Before the start of the study, a preliminary pre-test was applied to groups A and B, then the control group A was taught the natural numbers and integers in the traditional way using only the textbook, while the experimental group B was taught using puzzles and games in mathematics classes, in addition to the textbook. The study period lasted three weeks, 12 hours were given distributed over 18 class periods.
	
After the completion of the study, the mathematical thinking test was applied to the two groups as a post-test, and the data were analyzed using a t-test for two independent samples to test and judge the study hypotheses~\cite{Mendenhall}.

\subsection{The Mathematical Thinking Test}\label{sec6.2}
	
The mathematical thinking test that was applied in the current study was prepared by the researchers, and the test was corrected by assigning one score to each question, thus the total score of the mathematical thinking test was $12$. The mathematical thinking test was applied to a sample of size $40$ students from the study population and the consistency of each test item was calculated using Pearson's correlation between the item scores and the entire test scores.
	
The correlation coefficient was $0.855$ (see Table~\ref{tabn1}), so, there is a strong, positive, and highly statistically significant linear correlation, The p-value associated with this correlation is: $p \approx 2.28 \times 10^{-12} \approx 0.00000000000228$ $(r = 0.855,\ p < 0.001,\ n = 40)$ between the variables represented by the first two columns in the provided dataset. This confirms a robust and reliable linear association where increases in one variable are strongly associated with increases in the other.
	
\begin{table}[H]
\centering
\begin{tabular}{|c|c|c|c|c|c|} \hline
		\textbf{Sample 1} & \textbf{Sample 2} & \textbf{N} & \textbf{Correlation} & \textbf{95\% CI for $\boldsymbol{\rho}$} & \textbf{P-Value} \\ \hline
		C2 & C1 & 40 & 0.855 & (0.740, 0.921) & 0.001 \\ \hline
\end{tabular}
\caption{Pairwise Pearson Correlations}~\label{tabn1}
\end{table}

\section{Study results and analysis}\label{sec7}
	
This study aims to determine whether there is a statistically significant difference in test scores on natural numbers and integers between two groups of sixth-grade students: Group A (\textit{control}, traditional learning) and Group B (\textit{experimental}, math puzzles-based learning). We are interested in answering two specific questions about the impact of teaching method on mathematical thinking and mental arithmetic ability.
	
We start with Group~A, which had a size of $n=42$ with mean $\bar{x}_A = 65.357$ and standard deviation $s_A = 21.611$. Group~B had size $n=42$ with mean $\bar{x}_B = 77.833$ and standard deviation $s_B = 20.19$ (see individual sample statistics in Table~\ref{tab:descriptive}). With $p < 0.001$, we conclude the mean of Group~B is significantly greater than the mean of Group~A at $\alpha = 0.05$ significance level ($\bar{x}_B > \bar{x}_A$).
	
The critical $t$-value for a 95\% two-sided confidence interval with approximately $81.62$ degrees of freedom is $t^* \approx 1.989$. The margin of error is calculated as $1.989 \times 4.561 \approx 9.071.$ The two-sided 95\% confidence interval for the difference $(\mu_A - \mu_B)$ is $-12.476 \pm 9.071,$
resulting in an interval from $(-12.476 - 9.071)$ to $(-12.476 + 9.071)$, which is approximately $[-21.547, -3.405]$. This means we are 95\% confident that the true average score for Group~A (traditional learning) is between $3.405$ and $21.547$ points lower than Group~B (puzzle-based learning).
    
The 95\% lower bound for the difference in means $(\mu_A - \mu_B)$, derived from a two-sided confidence interval, is approximately $-21.547$. The descriptive statistics indicate that, on average, students in the experimental group (B) achieved higher scores on the test of natural numbers and integers compared to students in the control group (A). The variability in scores, as indicated by the standard deviations ($s_A = 21.611$, $s_B = 20.19$), was comparable between the two groups. We denote upper bound by $\operatorname{UB}$ and lower bound by $\operatorname{LB}$.
	
\begin{table}[H]
\centering
\begin{tabular}{|l|c|c|c|c|c|}
\hline
\textbf{Group} & \textbf{Sample Size (n)} & \textbf{Mean} & \textbf{SD} & \textbf{95\% $\operatorname{UB}$} & \textbf{95\% $\operatorname{LB}$} \\ \hline
A (Control) & 42 & 65.357 & 21.611 & 70.970 & 59.745 \\ \hline
B (Experimental) & 42 & 77.833 & 20.190 & 83.076 & 72.590 \\
\hline
\end{tabular}
\caption{Descriptive Statistics}~\label{tab:descriptive}
\end{table}
	
From Table~\ref{tab4}, this two-sided 95\% confidence interval:	The 95\% Lower Bound for the difference $(\mu_A - \mu_B)$ is approximately $-21.547$, 
	and the 95\% Upper Bound for the difference $(\mu_A - \mu_B)$ is approximately $-3.405$. 
	This means we are 95\% confident that the true average score for Group~A (traditional learning) 
	is at least $3.405$ points lower than Group~B (puzzle-based learning), and could be as much 
	as $21.547$ points lower. More specifically, the 95\% lower bound of $-21.547$ indicates that 
	the smallest plausible value for the true difference $(\mu_A - \mu_B)$ is $-21.547$, with 95\% confidence.
	
	The primary goal of this study was to determine if a statistically significant difference exists 
	between the average test scores of the control group (GA) and the experimental group (GB) at an 
	alpha level of $\alpha = 0.05$.
	
	To determine if the observed difference in mean scores is statistically significant, an independent 
	two-sample $t$-test (Welch's $t$-test, assuming unequal variances) was performed. The research 
	question about a ``statistically significant difference'' implies a two-tailed hypothesis test.
	
\begin{itemize}
\item \textbf{Null Hypothesis} ($H_0$): There is no statistically significant difference between the average scores of Group~A and Group~B ($\mu_A = \mu_B$).
		
\item \textbf{Alternative Hypothesis} ($H_1$): There is a statistically significant difference between the average scores of Group~A and Group~B ($\mu_A \neq \mu_B$).
\item \textbf{Significance Level} ($\alpha = 0.05$). The results of the $t$-test are: 
$t$-statistic $t = -2.7339$, degrees of freedom $df = 81.62$ and $p$-value $p \approx 0.0077$. The decision rule is: if the $p$-value obtained from the test is less than the chosen significance  level ($\alpha = 0.05$), the null hypothesis ($H_0$) is rejected in favor of the alternative hypothesis ($H_1$).
\end{itemize}

Comparing this $p$-value to the significance level: $p \approx 0.0077 < 0.05$, we reject the null hypothesis. This indicates that there is a statistically significant difference between the average scores of Group~A and Group~B. Specifically, the mean score for Group~B ($\bar{x}_B = 77.833$), which learned using math puzzles, 
is significantly higher than the mean score for Group~A ($\bar{x}_A = 65.357$), which learned through traditional methods. The difference in means is $\bar{x}_B - \bar{x}_A = 12.476$ points. 

The 95\% confidence interval for the difference between the means $(\mu_A - \mu_B)$, as suggested by Figure~\ref{fig1} (consistent with the $p$-value), would not contain zero, further supporting the significant difference. We conclude that the following homogeneity of variances:
\begin{enumerate}
    \item F-test ratio: $ F = \dfrac{s_A^2}{s_B^2} = \dfrac{467.02}{407.65} \approx 1.15 $.
    \item Critical $ F $-value ($\alpha = 0.05$, $ df_1 = 41 $, $ df_2 = 41 $) $\approx 1.69$.
    \item Conclusion: Variances are equal ($ F < 1.69 $); use pooled t-test.
\end{enumerate}

	\noindent By pooled variance we find 
\begin{equation}~\label{eq1}
s_p^2 = \frac{(n_A - 1)s_A^2 + (n_B - 1)s_B^2}{n_A + n_B - 2} .
\end{equation}
From~\eqref{eq1} we find that 
	\[
s_p^2	= \frac{41 \times 467.02 + 41 \times 407.65}{82} 
	\approx 437.34
	\]
Thus, $ s_p = \sqrt{437.34} \approx 20.91 $.
	
By using t-Statistic, we find that 
\begin{equation}~\label{eq2}
t = \frac{\bar{x}_B - \bar{x}_A}{s_p \sqrt{\dfrac{1}{n_A} + \dfrac{1}{n_B}}} . 
\end{equation}
Thus, from~\eqref{eq2} we find that
	\[
	t= \frac{77.833 - 65.357}{20.91 \times \sqrt{\dfrac{1}{42} + \dfrac{1}{42}}} 
	\approx 2.73
	\]
Noticed that: 
\begin{itemize}
    \item Degrees of Freedom $ df = n_A + n_B - 2 = 42 + 42 - 2 = 82 $.
    \item  Critical t-Value ($\alpha = 0.05$, two-tailed) $= \pm 1.989$.
\end{itemize}

The findings suggest that the teaching method employing common puzzles had a positive and statistically significant impact on students' performance on the test of natural numbers and integers. Assuming this test is a valid measure of mathematical thinking and mental arithmetic ability, it can be inferred that the puzzle-based learning approach contributed to an enhancement in these cognitive skills among the sixth-grade students in the experimental group.
	
Based on the independent two-sample t-test, there is a statistically significant difference $ (t(81.62) = -2.7339,\ p \approx 0.0077) $ at the significance level $ \alpha = 0.05 $ at the $\alpha=0.05$ level between the average scores of the control group (GA, traditional learning) and the experimental group (GB, puzzle-based learning) on the test of natural numbers and integers.
	
Students in the experimental group, who were taught using common puzzles, demonstrated significantly higher average scores compared to students in the control group. This suggests that the puzzle-based teaching methodology is more effective in improving student performance on tests measuring understanding of natural numbers and integers, and by extension, potentially enhancing mathematical thinking and mental arithmetic abilities as assessed by this test. Figure~\ref{fig1} as: 
\begin{figure}[H]
    \centering
		\includegraphics[width=0.60\linewidth]{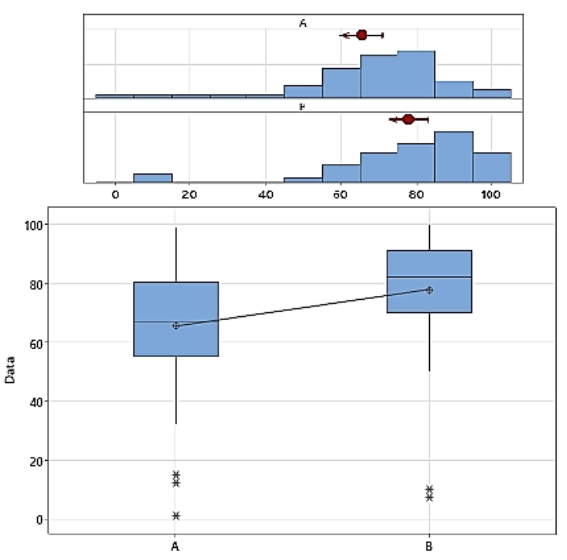}
		\caption{Distribution of Data and Boxplot of A, B.}
		\label{fig1}
\end{figure}
	
We can see with \emph{MINITAB} provided~\ref{fig1} above a one-sided p-value of $P=0.004$ for the hypothesis that the mean of A is less than the mean of B. Our calculations for a one-sided test (alternative='less') yielded a p-value of 0.0038. This confirms the direction of the difference (Group B scores higher than Group A) and its significance. The Power Analysis information in our study for a two-sample independent t-test ($\alpha=0.05$) and sample sizes of $42$ per group in the following Table 4 shows the power analysis:
	
	\begin{table}[H]
    \centering
		\begin{tabular}{|c|c|}
			\hline
			Detectable Difference (A - B) & Power \\
			\hline
			-9.47 & $60 \%$ \\
			\hline
			-10.82 & $70 \%$ \\
			\hline
			-12.40 & $80 \%$ \\
			\hline
			-14.60 & $90 \%$ \\
			\hline
			Observed Difference & -17.54 \\
			\hline
		\end{tabular}
		\caption{Power Analysis}\label{tab4}
	\end{table}
From the previous Table~\ref{tab4}, this observed difference of -12.476 falls between -11.442 (which has 80\% power) and -13.467 (which has 90\% power). This suggests that your study had a good statistical power (specifically, between 80\% and 90\%) to detect a true difference of the observed size. Power analysis helps us understand the probability of a study detecting an effect of a certain size, given the sample size and alpha level. Therefore, we conclude that the power to detect an observed difference is very good.

\section{Conclusion}\label{sec8}
	
The study contributes valuable knowledge to the field of mathematics education, especially concerning teaching in difficult circumstances and the potential of alternative pedagogical tools. We conducted a research study to evaluate the impact of using popular mathematical puzzles on the mathematical thinking of elementary school students.
	
The study compared the performance of two groups of sixth grade elementary school students in learning the natural numbers and integers: Group A (control group, taught through traditional instruction) and Group B (experimental group, taught through mathematical puzzles). Group B significantly outperformed Group A.  A paired experimental study (Pretest-Posttest Control Group Design) was conducted with the data we obtained to assess the effect of mathematical puzzles on students' problem-solving abilities, and then we measured the students' skills before they started solving mathematical puzzles and then after a period of solving the puzzles.
	
The results remained statistically and practically significant, suggesting that puzzle-based math learning enhances understanding and math skills more effectively than traditional methods and increases students' problem-solving abilities. These findings highlight the potential of innovative teaching strategies in math education and call for a broader exploration of interactive methods such as puzzles.
	
This study had a good power (between 80\% and 90\%), we can be more confident in our conclusion that the puzzle-based learning method (Group B) resulted in significantly higher scores. Therefore, the study's power emphasizes reliability, and the results call for the integration of puzzle-based methods in math education. The results showed a positive impact of puzzles, offer practical and relatively low-cost strategies for teachers working in resource-constrained and challenging environments. This is a significant plus for journals looking for research with actionable insights. Future research could explore the long-term effects of this teaching method and its applicability across different mathematical topics and age groups. Also, I suggest research with different age groups.


\begin{thebibliography}{99}
\bibitem{Kubinova} Kubinova, M., Novotna, J. \& Littler, G. H. (1998). ``Projects And Mathematical Puzzles-a Tool For Development Of Mathematical Thinking'', European Research in Mathematics Education, I.II, Group 5. \url{https://www.researchgate.net}.
\medskip

\bibitem{Borovik} Borovik, A. V. (2014). ``Calling a spade a spade: Mathematics in the new pattern of division of labour'', arXiv:1407.1954. \url{https://doi.org/10.48550/arXiv.1407.1954}.
\medskip
 
\bibitem{Mendenhall} Mendenhall, W., Beaver, R. J., \& Beaver, B. M. (2008). ``Introduction to probability and statistics'', 13th ed, USA:Cengage Learning.
 \url{ISBN	0495389536, 9780495389538}.
 \medskip

\bibitem{Almustafa} Almustafa, A., \& Alashkar, A. (2019). ``A Prospectus for Inclusion: Project study for a potential education solution for Syria’s Persons with Disability (PwDs)''. \url{http://dx.doi.org/10.31235/osf.io/n394v}.
\medskip

 \bibitem{Djordjevic} Djordjevic, G. S., Pavlovic-Babic, D., \& Stankovic, J. (2011). ``Evaluation Of High School Programme For Gifted Pupils In Physics And Sciences In Serbia-experience In Regional Cooperation-seenet-mtp Network''. \url{https://doi.org/10.48550/arXiv.1110.5072}.
 \medskip
 
\bibitem{Englert} Englert, C., Miller, D. J., \& Smaranda, D. D. (2020). ``The Weinberg Angle and 5D RGE effects in a SO(11) GUT theory''. \url{doi: 10.1016/j.physletb.2020.135548}.

\bibitem{Grcar} Grcar, J. F. (2011). ``Mathematics Turned Inside Out: The Intensive Faculty Versus the Extensive Faculty''. \url{https://doi: 10.1007/s10734-010-9358-y}.

\bibitem{Jorgensen} Jorgensen, P. E. T. (2006). Some recent trends from research mathematics and their connections to teaching: Case studies inspired by parallel developments in science and technology. \url{https://doi.org/10.48550/arXiv.math/0609209}.

\bibitem{Popel} Popel, M. (2014). ``The Methodical Aspects Of The Algebra and The Mathematical Analysis Study Using The Sagemath Cloud'', Information Technologies in Education, 19, 93–100. \url{doi: 10.14308/ite000488}.

\end{thebibliography}
\end{document}